\documentclass[10pt]{iopart}
\usepackage{setstack}
\usepackage{iopams}
\usepackage{amssymb}
\usepackage{graphicx}

\begin{document}

\title{Optomechanical sideband cooling of a thin membrane within a cavity}

\author{M. Karuza$^{1,2}$, C. Molinelli$^1$, M. Galassi$^1$, C. Biancofiore$^1$,R. Natali$^1$, P. Tombesi$^1$, G. Di Giuseppe$^1$, D. Vitali$^1$}
\address{$^1$ School of Science and Technology, Physics Division, University of Camerino, via Madonna delle Carceri, 9, I-62032 Camerino (MC), Italy, and INFN, Sezione di Perugia, Italy}
\address{$^2$ Department of Physics, University of Rijeka, Omladinska 14, HR-51000
Rijeka, Croatia}

\begin{abstract}
We present an experimental study of dynamical back-action cooling of the fundamental vibrational mode of a thin semitransparent membrane placed within a high-finesse optical cavity. We study how the radiation pressure interaction modifies the mechanical response of the vibrational mode, and the experimental results are in agreement with a Langevin equation description of the coupled dynamics. The experiments are carried out in the resolved sideband regime, and we have observed cooling by a factor $\sim 350$. We have also observed the mechanical frequency shift associated with the quadratic term in the expansion of the cavity mode frequency versus the effective membrane position, which is typically negligible in other cavity optomechanical devices.
\end{abstract}

\pacs{42.50.Wk, 42.50.Lc, 85.85.+j, 03.65.Ta}

\vspace{2pc}
\noindent{\it Keywords}: Mechanical effects of light, radiation pressure, cooling



\section{Introduction}

The preparation and manipulation of macroscopic mechanical systems in the quantum regime has received significant interest over the past decade~\cite{kippenberg,amo,marquardt-girvin,favero09}.
In fact, mechanical resonators provide unique opportunities in disparate field of applications, such as the detection of forces~\cite{VirgoNature}, displacements and masses~\cite{Roukes2009} at the ultimate limits imposed by the Heisenberg principle, the realization of quantum information architectures where they can act as universal quantum bus~\cite{Rabl2010}, and for fundamental tests of quantum theory~\cite{Marshall2003,Pikovski2012}. An important prerequisite for operating in the quantum regime is reducing as much as possible thermal noise effect, and this can be obtained by cooling the mechanical resonator close to its quantum ground state. Important results have been recently achieved in this respect. A GHz frequency piezomechanical oscillator has been cooled to the quantum regime with conventional cryogenics, and then probed by a superconducting qubit~\cite{O'Connell2010}. In contrast, cooling schemes based on dynamical back-action caused by the parametric coupling with an optical or microwave cavity~\cite{Braginski1967} can be applied to a wider class of nanomechanical and micromechanical
resonators which, due to the lower resonance frequency, are practically impossible to cool using only cryogenic techniques. In particular, it has been theoretically shown that dynamical back-action cooling allows reaching the quantum ground state of a mechanical mode~\cite{marquardt,wilson-rae,Genes08,dantan08} in the resolved sideband regime where the cavity linewidth is much smaller than the mechanical frequency. In this limit and if the cavity is resonant with the anti-Stokes sideband of the driving laser, effective phonon emission into the cavity is enhanced and phonon absorption is suppressed, yielding a large net laser cooling rate. After first demonstrations~\cite{gigan06,arcizet06,vahalacool,mavalvala,markus09,sideband}, sideband cooling has been recently employed to reach a phonon occupancy $n_{\rm eff}< 1$, either for an aluminium membrane capacitively coupled to a cryogenic microwave cavity~\cite{Teufel2011}, and for an integrated optical and mechanical nanoscale resonator realized in a photonic crystal structure~\cite{Chan2011}.

Dynamical back-action cooling can be equivalently described as a consequence of the modification of the mechanical susceptibility of the resonator caused by the optomechanical interaction. In fact, the back-action of a detuned cavity modifies both frequency (the so-called ``optical spring effect''~\cite{mavalvala,opticalspring}) and damping of the mechanical resonator. When the resonator is overdamped due to back-action, its susceptibility at resonance is strongly suppressed. As a consequence, the resonator becomes less sensitive to thermal noise, and this leads to cooling. The study of this modification of the mechanical response as a function of the cavity detuning has been carried out only in silica toroidal optomechanical systems~\cite{Riviere2011}. Here we perform a detailed study of dynamical back-action cooling and of radiation-pressure modifications of the mechanical properties in a membrane-in-the-middle (MIM) system formed by a vibrating thin silicon nitride (SiN) semitransparent membrane with high mechanical quality factor, placed within a high-finesse cavity. Such a scheme has been introduced in Ref.~\cite{harris} and has been then studied also in~\cite{njp,kimble2} (see also Ref.~\cite{polzik} for a similar scheme with a GaAs membrane cooled via an optical absorption process). We find that the experimental results are in excellent agreement with the theoretical prediction of Ref.~\cite{Biancofiore2011}, which included also the effect of membrane absorption and of an additional frequency shift caused by the presence of terms quadratic in the effective membrane position $\hat{q}$, which are typically negligible in most optomechanical systems. We measure the effective temperature of the cooled vibrational mode in three different ways, obtaining consistent results, in agreement with the theoretical expectations. In particular we demonstrate resolved sideband cooling by a factor $\sim 350$ starting from room temperature.

The paper is organized as follows. Sec.~II describes the experimental setup, Sec.~III adapts the theoretical description of Ref.~\cite{Biancofiore2011} to the present experimental conditions. Sec.~IV illustrates the experimental results and their matching with the theory predictions for the mechanical frequency shift, the effective damping and the effective temperature. Sec.~V is for concluding remarks.

\section{The experimental setup}\label{sec:setup}

Our MIM setup is schematically described in Fig.~1. Laser light at $\lambda = 1064$ nm is produced by a Nd:YAG laser
(Innolight). After exiting the laser head, the light beam encounters a half wave plate (HWP) that is followed by a polarizing beam splitter
(PBS). By rotating the HWP the optical power can be distributed between a probe and a pump beam, with frequencies $\omega_p$ and $\omega_L$ respectively. During measurements the probe beam power was 100 $\mu$W, while the rest, about 200 mW, went into the pump beam optical line where at the end only a small fraction was used. The detuning from the probe beam was
controlled by two cascaded acusto-optical modulators (AOMs) whose central operating frequency is 80 MHz. By selecting diffracted beams of
the first order but with opposite signs, detunings from 0 to 40 Mhz can be obtained, although only detunings up to 500 kHz have been used. The pump
beam intensity was controlled by the modulation amplitude of the electrical signal used to drive AOM$_2$ (see Fig.~1). After the AOMs the pump beam passes
through an optical isolator (OFR$_2$) and is mode-matched to the Fabry-Perot (FP) optical cavity by means of two lenses (L$_1$ and L$_3$).
Before being injected in the FP cavity the pump beam is then combined with the probe
beam by means of PBS$_2$.

The cavity is $L \simeq 93$ mm long and consists of two equal dielectric mirrors, each with a radius of curvature $R = 10$ cm. The measured value of the empty cavity finesse is $F \simeq 60000$ and is consistent with the mirror's nominal reflectivity. Halfway between the mirrors a thin stoichiometric silicon nitride membrane is mounted on series of piezo-motor driven optical mounts that control the angular alignment as well as the linear positioning with respect to the optical axis.

The membrane is a commercial 1 mm $\times$ 1 mm ${\rm Si_3 N_4}$ stoichiometric x-ray window (Norcada), with nominal thickness $L_d =50$ nm, and index of refraction $n_R \simeq 2$, supported on a 200 $\mu$m Si frame. It has been chosen due to its high mechanical quality factor and very low optical absorption at $\lambda = 1064$ nm~\cite{Zwickl2008}. Its optical properties were also experimentally verified, yielding an intensity reflection coefficient equal to $R \simeq 0.18$, and an imaginary part of the index of refraction equal to $n_I \simeq 2 \times 10^{-6}$.

\begin{figure}[ht]
\includegraphics[width=0.8\textwidth]{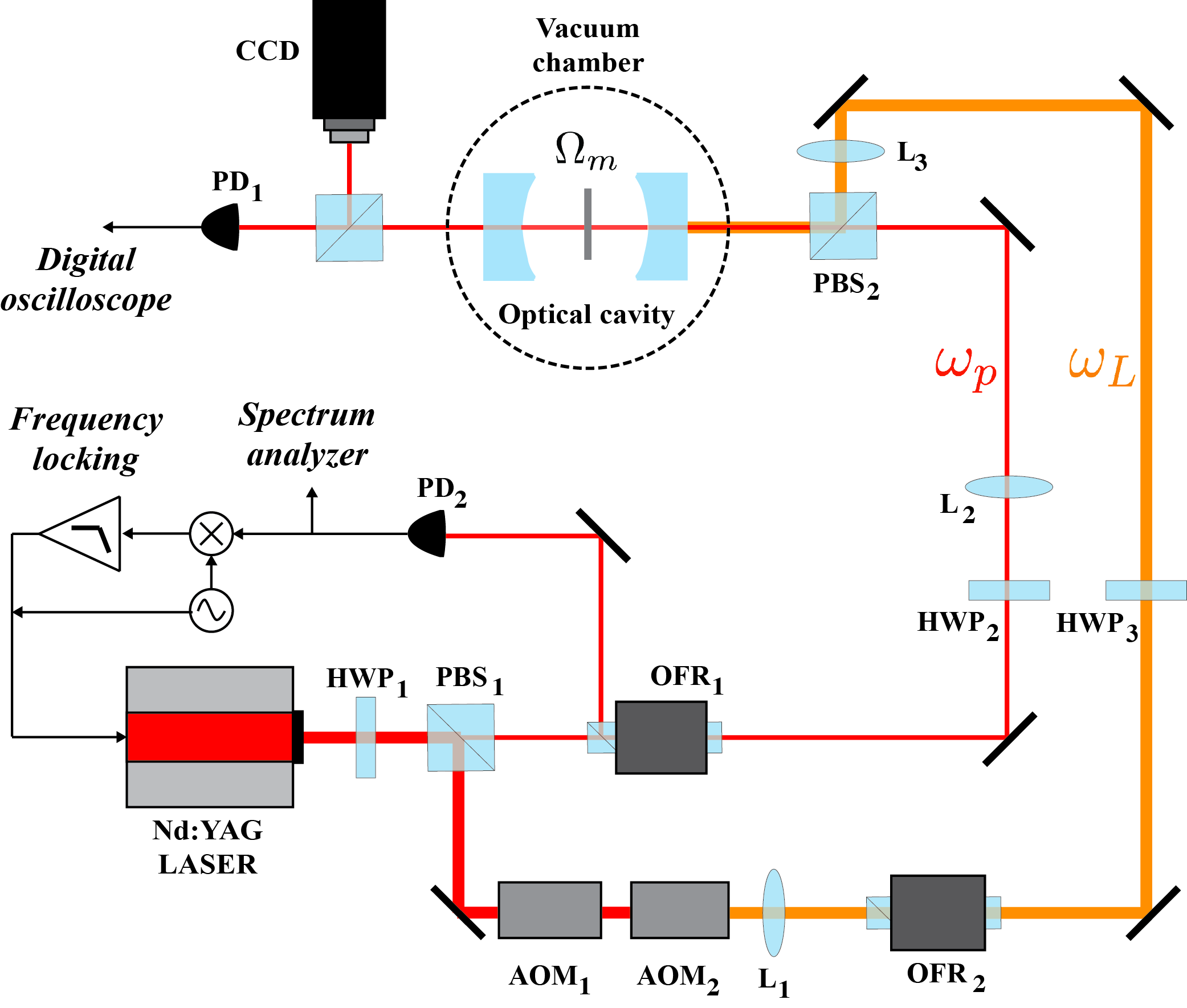}
\caption{Schematic description of the experimental setup.
 }
\label{fig:1}
\end{figure}

The membrane's mechanical motion is monitored by the probe beam with frequency $\omega_p$ that is also used for locking the laser to the FP instantaneous resonant frequency. This is done by observing the light reflected from the cavity by a photodiode PD$_2$, whose output signal is amplified and fed into a frequency locking loop and a spectrum analyzer where the membrane's mechanical motion is observed. The locking scheme is a standard Pound-Drever-Hall (PDH) scheme~\cite{Pound1946,Drever1983}, where the sidebands necessary for obtaining the error signal feeded into the locking loop are created by modulating directly the laser crystal~\cite{Bregant2002}. In order to avoid the deterioration of the mechanical properties of the membrane and optical properties of the FP cavity, the mirrors and membrane are mounted inside a vacuum chamber which is evacuated by a turbo-molecular pump down to $10^{-5}$ mbar. Once the base pressure is reached, the pump is switched off and disconnected from the vacuum chamber in order to minimize the mechanical noise. Due to some leaks, the residual pressure inside the chamber reaches values as high as $10^{-2}$ mbar in about one day. Although some worsening of the membrane's mechanical quality factor has been observed, it does not limit our sensitivity. A vibrationless pumping scheme is being prepared for future measurements.

\section{Quantum Langevin description}

Radiation pressure of the intracavity field excites the membrane vibrational modes, and therefore one has a multimode bosonic system in which mechanical and optical modes interact in a nonlinear way. However, one can adopt a simplified description based on a \emph{single} cavity mode interacting with a \emph{single} mechanical mode~\cite{harris,njp,kimble2,Biancofiore2011}. One can restrict to a single cavity mode if the driving laser populates a given cavity mode only (here a TEM$_{00}$ mode, associated with the annihilation operator $\hat{a}$), and if scattering into other modes is negligible~\cite{law}. Moreover, one can consider a single mechanical mode of the membrane (described by dimensionless position $\hat{q}$ and momentum $\hat{p}$ operators, such that $[\hat{q},\hat{p}]= i$) when the detection bandwidth is chosen so that it includes only a single, isolated, mechanical resonance with frequency $\Omega_m$.

By explicitly including cavity driving by the pump laser with frequency $\omega_L$ and input power ${\mathcal P}$, the system Hamiltonian reads
\begin{equation}
H=\frac{\hbar \Omega _{m}}{2}(\hat{p}^{2}+\hat{q}^{2})+\hbar \omega(\hat{q})\hat{a}^{\dagger }\hat{a}
+i\hbar E(\hat{a}^{\dagger }e^{i\omega_L t}-\hat{a}e^{-i\omega_L t}),  \label{eq:Ham-optomech}
\end{equation}
where $E=\sqrt{2\mathcal{P}\kappa_0 /\hbar \omega _{L}}$, with $\kappa_0$ the coupling rate through the input mirror.
The optomechanical interaction is described by the position-dependent optical frequency
\begin{equation}\label{eq:pos-dep-freq-gen}
\omega(\hat{q})=\omega_0+{\rm Re}\left\{\delta\omega\left[z_0(\hat{q})\right]\right\},
\end{equation}
where $\omega_0$ is the cavity mode frequency in the absence of the membrane, and ${\rm Re}\left\{\delta\omega\left[z_0(\hat{q})\right]\right\}$ is the frequency shift caused by the insertion of the membrane. This shift depends upon the membrane position along the cavity axis $z_0(\hat{q})$, which in turn depends upon the coordinate $\hat{q}$ because one can write $z_0(\hat{q}) = z_0+x_0 \Theta \hat{q}$, where $z_0$ is the membrane center-of-mass position along the cavity axis, $\Theta$ is the transverse overlap integral between the optical mode and the vibrational mode~\cite{Biancofiore2011}, and $x_0=\sqrt{\hbar/m \Omega_m}$, with $m$ the effective mass of the mechanical mode.

The system is also affected by fluctuation-dissipation processes; the mechanical mode undergoes a viscous force with damping rate $\gamma_m$ and a Brownian stochastic force with zero mean value $\hat{\xi}(t)$. We operate at room temperature $T$, and the correlation function of this Brownian noise is well approximated by~\cite{Landau,gard,GIOV01}
\begin{equation}
\label{eq:browncorre2}\left\langle \hat{\xi}(t)\hat{\xi}(t^{\prime})\right\rangle \simeq \gamma_{m}\left[(2n+1) \delta(t-t^{\prime})+i
\frac{\delta^{\prime}(t-t^{\prime})}{\Omega_{m}}\right]  ,
\end{equation}
where $k_B$ is the Boltzmann constant, $n=\left[\exp (\hbar \Omega _{m}/k_{B}T)-1\right] ^{-1} \simeq k_B T/\hbar \Omega_m$ is the mean thermal phonon number at $T$, and $\delta^{\prime}(t-t^{\prime})$ denotes the derivative of the Dirac delta.

The cavity mode looses photons through the input mirror (with decay rate $\kappa_0 $), the back mirror (with decay rate $\kappa_2 $), and also due to optical absorption by the membrane, with decay rate $\kappa_1(\hat{q})\equiv \left|{\rm Im}\left\{\delta\omega\left[z_0(\hat{q})\right]\right\}\right|$, which is nonzero owing to the small but nonzero imaginary part of the refraction index, $ n_I$~\cite{Biancofiore2011}. Optical absorption by the membrane depends upon the mechanical position operator $\hat{q}$, being therefore a nonlinear dissipative process affecting both the optical and the mechanical mode. Each decay channel is associated with a vacuum optical input noise $\hat{a}_j^{\rm in}(t)$, $j=0,1,2$, with correlation functions~\cite{gard}
\begin{equation}\label{eq:voinoise}
 \langle \hat{a}_k^{\rm in}(t)\hat{a}_j^{\rm in,\dag }(t^{\prime })\rangle = \delta_{kj}\delta (t-t^{\prime }).
\end{equation}
Adding the above damping and noise terms to the Heisenberg equations of motion derived from the Hamiltonian of Eq.~(\ref{eq:Ham-optomech}), one gets the following set of nonlinear quantum Langevin equations (QLE) which, in the frame rotating at the pump laser frequency $\omega_{L}$, read
\begin{eqnarray}
\dot{\hat{q}}& =& \Omega _{m}\hat{p}, \label{nonlinlang1}\\
\dot{\hat{p}}& = & -\Omega _{m}\hat{q}-\gamma _{m}\hat{p}-\partial_q \omega(\hat{q}) \hat{a}^{\dagger }\hat{a}+\hat{\xi} -i\frac{\partial_q \kappa_1(\hat{q})}{\sqrt {2\kappa_1(\hat{q})}}\left[\hat{a}^{\dagger}\hat{a}_1^{\rm in}-\hat{a} \hat{a}_1^{\rm in,\dagger}\right], \label{nonlinlang2}\\
\dot{\hat{a}}& = & -i\left[\omega(\hat{q})-\omega_L\right]\hat{a}-\kappa_T(\hat{q})\hat{a}
+E \nonumber \\
&& +\sqrt{2\kappa_0}\hat{a}_0^{\rm in}+\sqrt{2\kappa_1(\hat{q})}\hat{a}_1^{\rm in}+\sqrt{2\kappa_2}\hat{a}_2^{\rm in}, \label{nonlinlang3}
\end{eqnarray}
where $\partial_q$ denotes the derivative with respect to $\hat{q}$, and $\kappa_T(\hat{q})=\kappa_0+\kappa_1(\hat{q})+\kappa_2$ is the total cavity decay rate.

Eqs.~(\ref{nonlinlang1})-(\ref{nonlinlang3}) illustrate the peculiar aspects of the MIM scheme with respect to the paradigm optomechanical system represented by a Fabry-Perot cavity with a highly reflecting movable micro-mirror, which satisfactorily applies to a large number of optomechanical devices~\cite{Aspelmeyer2010}. In this latter scheme $\omega(\hat{q})=\omega_0(1-\hat{q}/L)$ and optical absorption can be usually neglected ($\kappa_1(\hat{q})\simeq 0$), and therefore the nonlinearities in $\hat{q}$ appearing in Eqs.~(\ref{nonlinlang1})-(\ref{nonlinlang3}) are absent. Our experimental results will show that the quadratic term in the power expansion of $\omega(\hat{q})$ has appreciable effects on the optically induced mechanical frequency shift. On the contrary we will see that membrane absorption can be neglected also in our case, due to the very low value of $n_I$ of the stoichiometric Si$_3$N$_4$ membrane employed here.

\subsection{Linearized quantum Langevin equations}

Our experiment is carried out in the usual ``linearized'' regime characterized by an intense stationary intracavity field with amplitude $\alpha_s$ ($|\alpha_s| \gg 1$), easily achievable with moderate pump power due to the large cavity finesse $F$. The vibrational mode of interest is correspondingly deformed, with a new stationary position $q_s$, satisfying, together with $\alpha_s$, the coupled nonlinear conditions
\begin{eqnarray}
\label{eq:stat1}
  q_s &=& -\frac{\partial_q \omega(q_s)|\alpha_s|^2}{\Omega_m}, \\
  |\alpha_s|^2 &=& \frac{E^2}{ \kappa_T(q_s)^2+\left[\omega_L-\omega(q_s)\right]^2}, \label{eq:stat2}
\end{eqnarray}
which may show optical bistability~\cite{dorsel,brennecke,Karuza2011}.

When the system is stable, the relevant dynamics concern the fluctuations of the cavity and mechanical modes around the classical steady-state described by Eqs.~(\ref{eq:stat1})-(\ref{eq:stat2}). Rewriting each Heisenberg operator of Eqs.~(\ref{nonlinlang1})-(\ref{nonlinlang3}) as the classical steady state value plus an additional
fluctuation operator with zero mean value, and neglecting all the nonlinear terms in the equations, one gets the following linearized QLE for the fluctuations~\cite{Biancofiore2011}
\begin{eqnarray}
\delta \dot{\hat{q}}& =&\Omega _{m}\delta \hat{p}, \label{lle1}\\
\delta \dot{\hat{p}}& =&-\left[\Omega _{m}+\partial_q^2 \omega(q_s)|\alpha_s|^2\right]\delta \hat{q}-\gamma _{m}\delta \hat{p}+\frac{G}{\sqrt{2}}\left(\delta \hat{a} +\delta \hat{a}^{\dagger}\right) \nonumber\\
&& +\hat{\xi} -i\frac{\Gamma}{2\sqrt {\kappa_1(q_s)}}\left(\hat{a}_1^{\rm in}-\hat{a}_1^{\rm in,\dag }\right), \label{lle2}\\
\delta \dot{\hat{a}}& =&-\left[\kappa_T(q_s)+i \Delta \right] \delta \hat{a}+\frac{i G -\Gamma }{\sqrt{2}}\delta \hat{q} \nonumber \\
&&+
\sqrt{2\kappa_0}\hat{a}_0^{\rm in}+\sqrt{2\kappa_1(q_s)}\hat{a}_1^{\rm in}+\sqrt{2\kappa_2}\hat{a}_2^{\rm in}. \label{lle3}
\end{eqnarray}
We have redefined the phase reference of the cavity field so that $\alpha _{s}$ is real and positive, and we have defined the effective detuning $\Delta = \omega(q_s)-\omega_L$ and $\Gamma=\sqrt{2}\partial_q \kappa_1(q_s)\alpha_s$. The linearized QLE of Eqs.~(\ref{lle1})-(\ref{lle3}) show that the mechanical and cavity mode
fluctuations are coupled by the effective optomechanical coupling
\begin{equation}\label{optoc}
G = -\sqrt{2}\partial_q \omega(q_s)\alpha_{s}  = -2\left(\frac{\partial\omega}{\partial z_0}\right)\Theta \sqrt{\frac{\mathcal{P}\kappa_0 }{m \Omega_m \omega _{L}\left[\kappa_T(q_s)^{2}+\Delta ^{2}\right] }}, 
\end{equation}
which can be made large by increasing the intracavity amplitude $\alpha _{s}$. Furthermore $G$ can be fine-tuned in the MIM system by shifting the membrane along the cavity axis, thereby changing $\partial \omega/\partial z_0$.

\section{Effect of radiation pressure on the membrane vibrational mode}

We observe the motion of the vibrational mode of interest by detecting the noise spectrum of the phase of the resonant weak probe field at $\omega_p$ reflected by the cavity, $\phi(\omega)$. This phase is related to the effective position $\delta \hat{q}$ by
\begin{equation}
\phi(\omega)= \frac{G_p}{\kappa_T(q_s)-i\omega} \delta \hat{q}(\omega)+s(\omega),
\label{detect}
\end{equation}
where
\begin{equation}
G_p = -2\left(\frac{\partial\omega}{\partial z_0}\right)\Theta_p \sqrt{\frac{\mathcal{P}_p\kappa_p }{m \Omega_m \omega _{p}\kappa_T(q_s)^{2} }},
\end{equation}
is the effective optomechanical coupling of the resonant probe beam, analogous to that of the driving field of Eq.~(\ref{optoc}), but with the corresponding input power $\mathcal{P}_p$, coupling rate $\kappa_p$, and overlap integral $\Theta_p$. $s(\omega)$ denotes the detection noise, essentially given by shot noise, which at resonance is uncorrelated with $\delta \hat{q}$, implying that the phase noise spectrum is given by
\begin{equation}
S_{\phi}(\omega)= \left[\frac{G_p^2}{\kappa_T(q_s)^2+\omega^2}\right] S_{q}(\omega)+S_s,
\label{detect2}
\end{equation}
where $S_{q}(\omega)$ is the spectrum of the dimensionless position $\hat{q}$, and $S_s$ is the (typically flat) shot noise spectrum. After calibration in m$^2$/Hz (and using $\hat{x}\equiv x_0 \hat{q}$), the detected spectrum can be written as
\begin{equation}
S_{x}^{\rm det}(\omega)= x_0^2 S_{q}(\omega)+ \left[\frac{\kappa_T(q_s)^2+\omega^2}{G_p^2}\right] x_0^2 S_s.
\label{detect3}
\end{equation}
The explicit expression of $S_{q}(\omega)$ is obtained by solving the linearized QLE of Eq.~(\ref{lle1})-(\ref{lle3}) in the frequency domain \cite{Genes08,output,Biancofiore2011} and reads
\begin{equation} \label{posspe}
S_{q}(\omega )=\left|\chi _{\rm eff}(\omega )\right|^{2}[S_{\rm th}(\omega )+S_{\rm rp}(\omega)+S_{\rm abs}(\omega)],
\end{equation}
with
\begin{equation}
S_{\rm th}(\omega )=\frac{\gamma _{m}\omega }{\Omega _{m}}\coth \left( \frac{%
\hbar \omega }{2k_{B}T}\right) \simeq  \frac{2\gamma _{m}k_B T}{\hbar \Omega_m}  \label{spectratherm}
\end{equation}%
the thermal noise spectrum,
\begin{equation}
S_{\rm rp}(\omega)=\frac{G^{2}\kappa_T(q_s) \left[ \Delta ^{2}+\kappa_T^{2}(q_s)
+\omega ^{2}\right] }{\left[ \kappa_T^{2}(q_s)+(\omega -\Delta )^{2}\right] %
\left[ \kappa_T^{2}(q_s)+(\omega +\Delta )^{2}\right] }  \label{spectrarpn}
\end{equation}%
the radiation pressure noise spectrum due to the intense pump beam at frequency $\omega_L$, and
\begin{equation}
S_{\rm abs}(\omega)=\frac{\Gamma^2}{4\kappa_1(q_s)}+\frac{\Gamma G \Delta \left[ \Delta ^{2}+\kappa_T^{2}(q_s)-\omega ^{2}\right] }{\left[ \kappa_T^{2}(q_s)+(\omega -\Delta )^{2}\right] %
\left[ \kappa_T^{2}(q_s)+(\omega +\Delta )^{2}\right] }  \label{spectraabs}
\end{equation}
the additional noise spectrum associated with membrane absorption~\cite{Biancofiore2011}.
The main effect of the optomechanical interaction on the membrane vibrational mode is the modification of its mechanical susceptibility (see Eq.~(\ref{posspe})), which becomes
\begin{equation}
\chi _{\rm eff}(\omega )=\frac{\Omega _{m}}{\tilde{\Omega}_{m}^{2}-\omega^{2}-i\omega \gamma _{m}-\frac{G\Omega _{m}\left[G\Delta-\Gamma\left[\kappa_T(q_s) -i\omega\right]\right]}{\left[\kappa_T(q_s) -i\omega
\right]^{2}+\Delta ^{2}}} ,\label{chieffD}
\end{equation}%
where $\tilde{\Omega}_{m}^{2}=\Omega_{m}^{2}+h \Omega_{m}$, with $h=\partial_q^2 \omega(q_s)|\alpha_s|^2=(\partial^2 \omega/\partial z_0^2)x_0^2 \Theta^2 |\alpha_s|^2$. This effective susceptibility can be read as the susceptibility of an oscillator
with effective (frequency-dependent) resonance frequency and damping rate respectively given by
\begin{eqnarray}
\Omega_{m}^{\rm eff}(\omega )&=&\left[ \tilde{\Omega} _{m}^{2}-\frac{G \Omega_{m}\left\{G \Delta \left[\kappa_T^{2}(q_s)-\omega ^{2}+\Delta ^{2}\right]-\Gamma\kappa_T(q_s)\left[\kappa_T^{2}(q_s)+\omega ^{2}+\Delta ^{2}\right]\right\}}{\left[ \kappa_T^{2}(q_s)+(\omega -\Delta )^{2}\right] %
\left[ \kappa_T^{2}(q_s)+(\omega +\Delta )^{2}\right] }\right]^{\frac{1}{2}},  \label{omegeff} \\
\gamma _{m}^{\rm eff}(\omega )&=&\gamma _{m}+\frac{G \Omega_{m}\left\{2G \Delta \kappa_T(q_s)-\Gamma\left[\kappa_T^{2}(q_s)+\omega ^{2}-\Delta ^{2}\right]\right\}}{\left[ \kappa_T^{2}(q_s)+(\omega -\Delta )^{2}\right] %
\left[ \kappa_T^{2}(q_s)+(\omega +\Delta )^{2}\right] }\label{dampeff}.
\end{eqnarray}%
The present experiment is carried out in the weak-coupling regime $G \ll \Omega_m $, and consequently the effective susceptibility $ \chi _{\rm eff}(\omega )$, even if appreciably modified, remains peaked approximately around $\Omega_m$. Therefore, the effective mechanical frequency and damping modified by the optomechanical interaction can be satisfactorily estimated by evaluating Eqs.~(\ref{omegeff}) and (\ref{dampeff}) at $\omega=\Omega_m$.

\begin{figure}[ht]
   \centering
  \includegraphics[width=.8\textwidth]{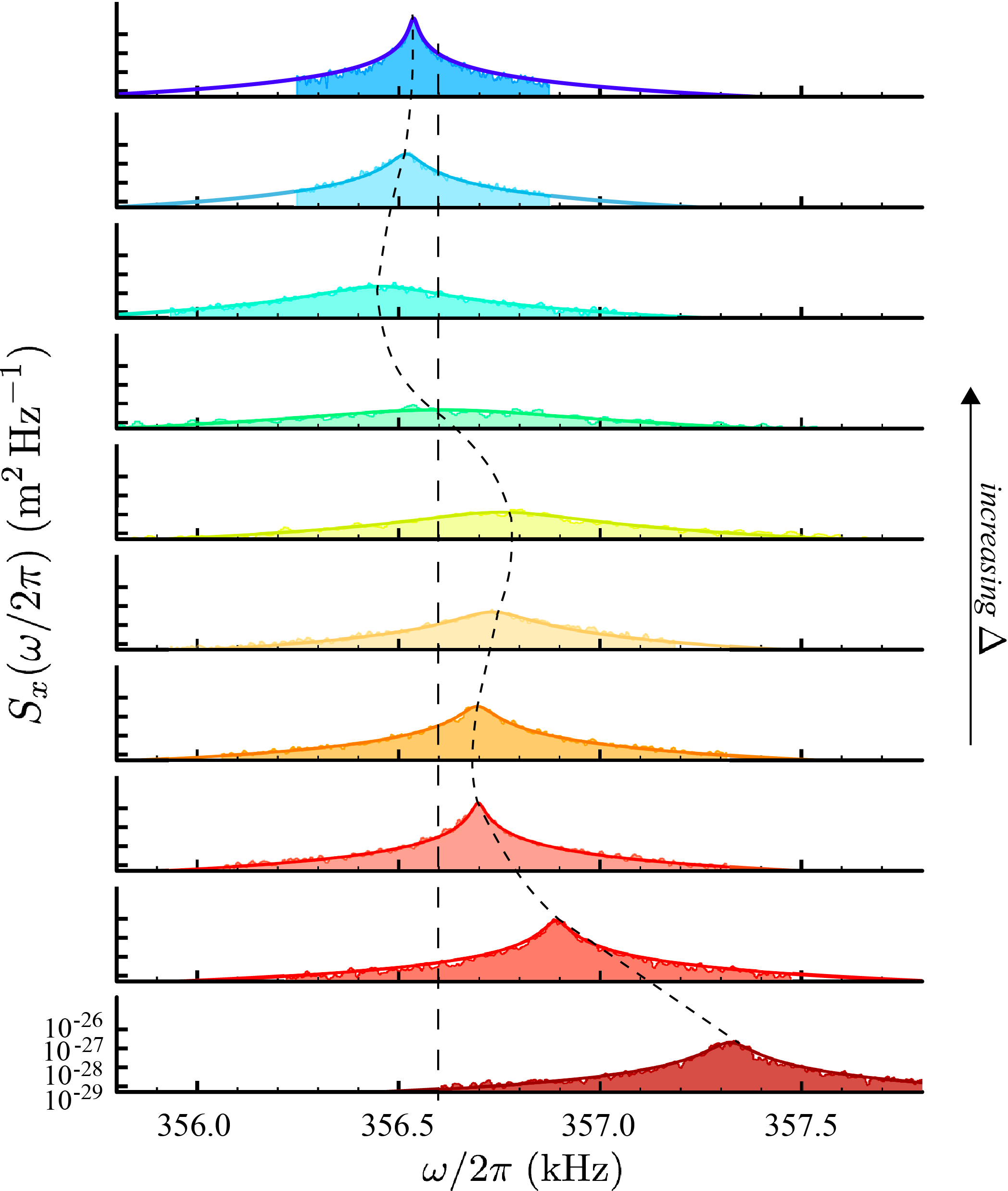}
   \caption{Calibrated position noise spectrum around the resonance associated with the fundamental vibrational mode of the membrane, with bare mechanical frequency $\Omega_m = 356.6$ kHz, mass $m=45$ ng, quality factor $Q=24000$, at different values of the detuning, $\Delta =30$, $60$, $180$, $280$, $320$, $340$, $355$, $380$, $410$, $600$ kHz, from the lower to the upper curve. The membrane is fixed at $10$ nm distance from a field node and the driving input power is ${\mathcal P}=670$ $\mu$W.}
   \label{fig:spectra}
\end{figure}

The modification of the mechanical response by the interaction with the driven cavity mode is clearly visible in Fig.~\ref{fig:spectra}, where the calibrated position spectrum is shown at different (positive) values of the detuning of the driving field $\Delta$: the mechanical resonance peak shifts and broadens by varying $\Delta$. We notice that just outside the resonance peak, the displacement sensitivity of our optomechanical detection system is equal to $\sqrt{S_x}\simeq 2.4 \times 10^{-15}$ m$/\sqrt{\rm Hz}\simeq 40 \sqrt{S_{x,{\rm SQL}}(\Omega_m)}$, where $S_{x,{\rm SQL}}(\Omega_m)=2\hbar Q/m\Omega_m^2$ is the standard quantum limit (SQL) for a displacement spectral measurement at its peak value at resonance. This value is not far from the value $\sqrt{S_x}=1.9 \times 10^{-16}$ m$/\sqrt{\rm Hz}$ recently achieved in a Michelson-Sagnac interferometer also based on a SiN membrane~\cite{Kaufer2012}.

\begin{figure}[h]
   \centering
   \includegraphics[width=.8\textwidth]{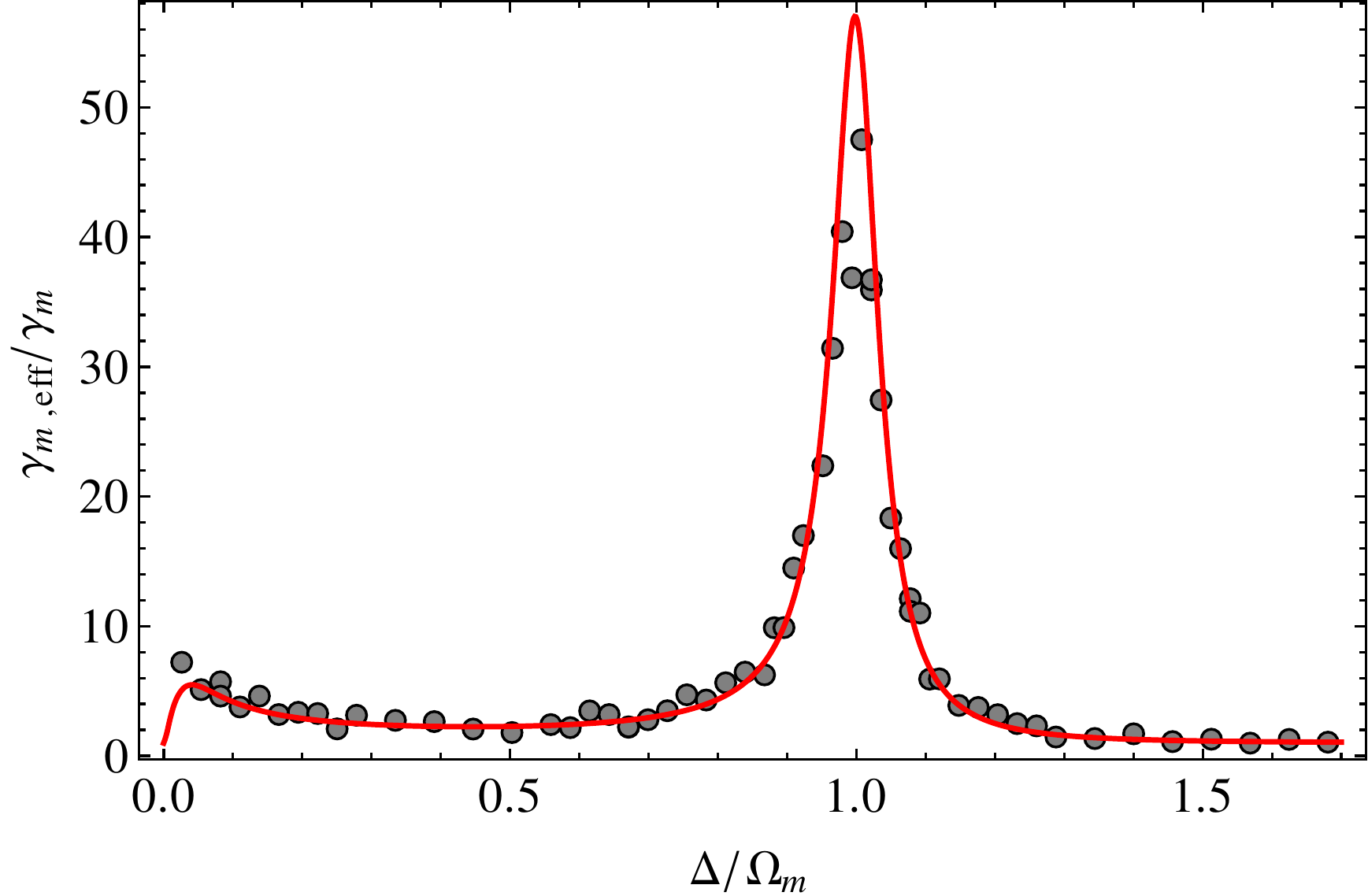}
   \caption{Scaled effective mechanical damping $\gamma_m^{\rm eff}/\gamma_m$ vs the scaled cavity-driving detuning $\Delta/\Omega_m$, in the region of positive detunings corresponding to a driving red-detuned with respect to the cavity mode. Dots are the experimentally measured values while the red full line refer to the prediction of Eq.~(\protect\ref{dampeff}). See Fig.~\protect\ref{fig:spectra} and text for the other parameter values.}
   \label{fig:dampeff}
\end{figure}

\begin{figure}[h]
   \centering
   \includegraphics[width=.8\textwidth]{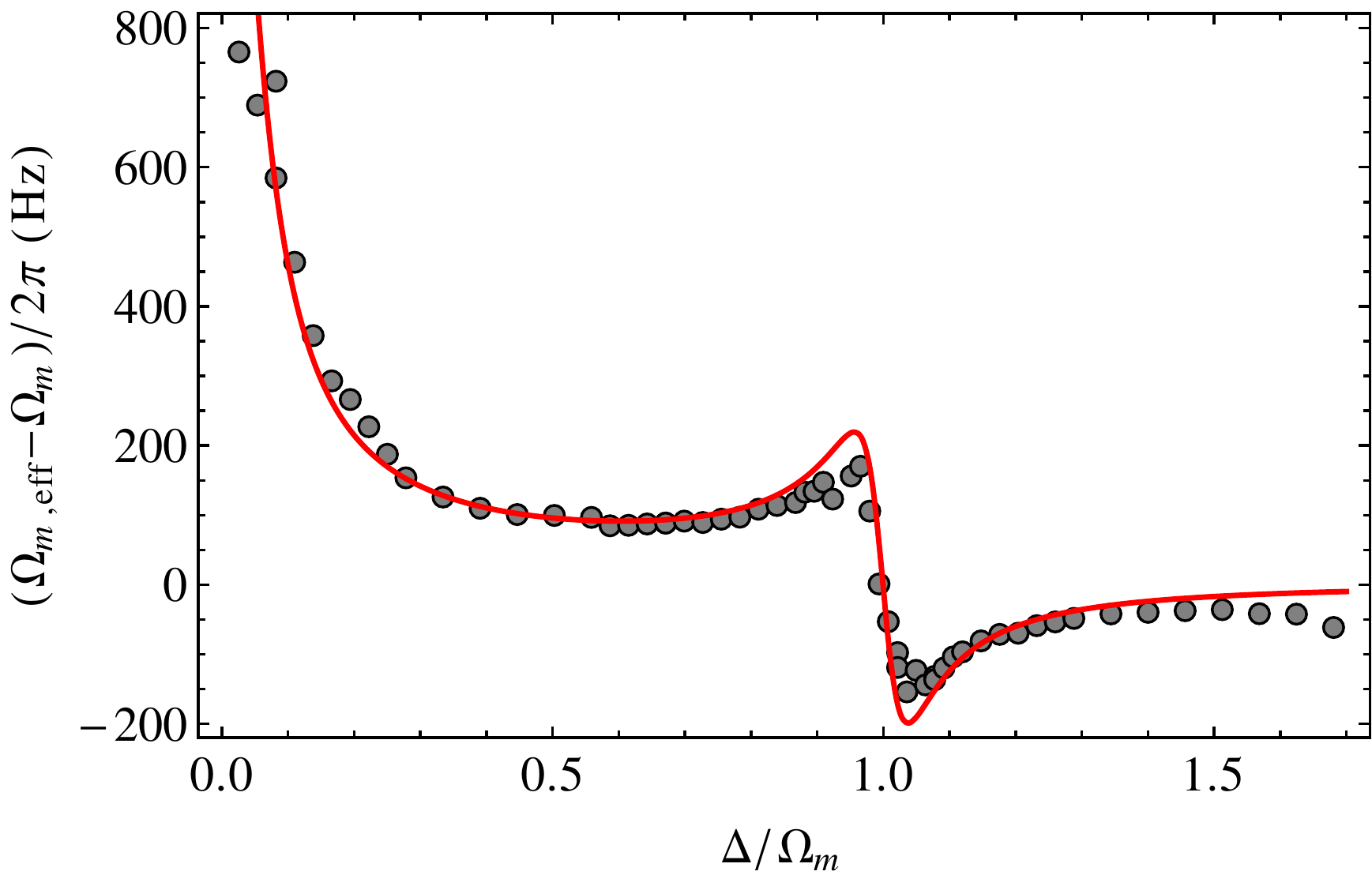}
   \caption{Mechanical frequency shift $\Omega_m^{\rm eff}-\Omega_m$ vs the scaled cavity-driving detuning $\Delta/\Omega_m$, in the region of positive detunings corresponding to a driving red-detuned with respect to the cavity mode. Dots are the experimentally measured values and the red full line refer to the prediction of Eq.~(\protect\ref{omegeff}). See Fig.~\protect\ref{fig:spectra} and text for the other parameter values. }
   \label{fig:omegeff}
\end{figure}

We have performed a detailed study of the dependence of the effective mechanical frequency $\Omega_m^{\rm eff}$ and effective mechanical damping $\gamma_m^{\rm eff}$ as a function of the detuning $\Delta$, which is shown in Figs.~\ref{fig:dampeff}-\ref{fig:omegeff}. Dots correspond to the experimental data obtained by a Lorentzian fit of the resonance peak in the detected spectrum, while the continuous red curves correspond to the theoretical prediction of Eqs.~(\ref{omegeff})-(\ref{dampeff}) (evaluated at $\omega=\Omega_m$), with the experimental parameter values $\Omega_m = 356.6$ kHz, mass $m=45$ ng, quality factor $Q=24000$, membrane at $10$ nm distance from a field node, overlap integral $\Theta \simeq 1$, driving input power ${\mathcal P}=670$ $\mu$W, and $\kappa_T(q_s)=77$ kHz. As expected, we see a significant increase of mechanical damping (and therefore cooling) at the resonant condition $\Delta=\Omega_m$, corresponding to the drive resonant with the red motional sideband of the cavity. In this condition, scattering of anti-Stokes (Stokes) photons into the cavity is enhanced (suppressed)~\cite{marquardt,wilson-rae,Genes08} and the net laser cooling rate is maximum.

The best-fit curve also yields a negligible value for the membrane absorption-related parameter $\Gamma$, $|\Gamma |\lesssim 4 \times 10^{-8} \Omega_m$ which is consistent with the value $n_I \simeq 2 \times 10^{-6}$ estimated in Ref.~\cite{Karuza2012} and reasonable for stoichiometric Si$_3$N$_4$ membranes~\cite{kimble2}. The best-fit curve instead yields a small, but non-negligible value for the parameter $h$ related with the mechanical frequency shift caused by the second-order term in the expansion of the position dependent cavity frequency $\omega(\hat{q})$, $h= 10^{-5} \Omega_m$ (around $\Delta=\Omega_m$).

\begin{figure}[h]
   \centering
   \includegraphics[width=.8\textwidth]{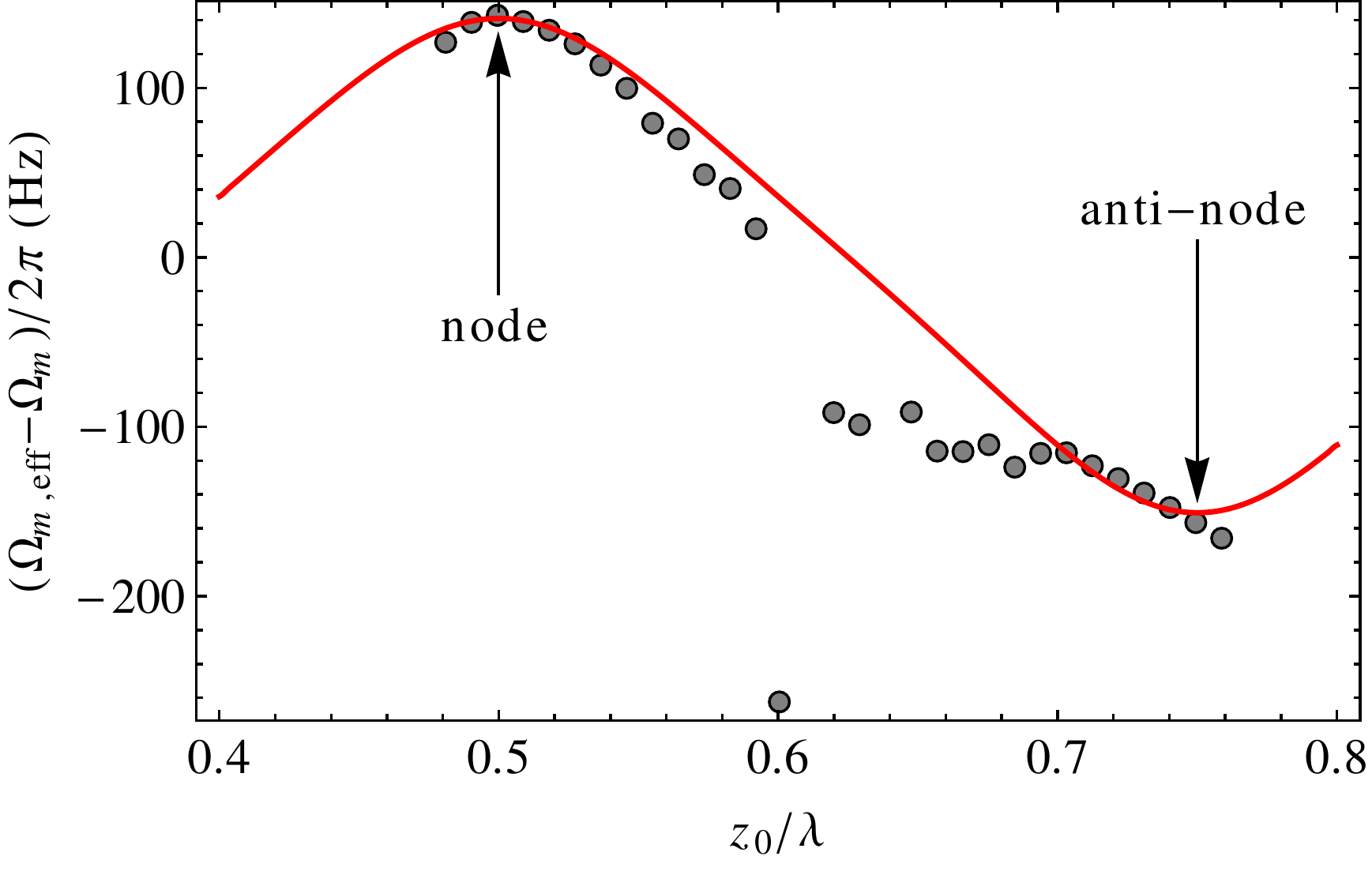}
   \caption{Mechanical frequency shift $\Omega_m^{\rm eff}-\Omega_m$ vs the membrane position along the cavity axis, from a node to an antinode of the cavity field, at the optical resonance $\Delta=0$. Dots refer to the measured values, the continuous red line refer to the prediction of Eq.~(\protect\ref{omegeff}). The other parameter values are as in Fig.~\protect\ref{fig:spectra} except that ${\mathcal P}=76$ $\mu$W. The narrow interval where the data depart from theory prediction corresponds to the presence of an avoided crossing between the driven TEM$_{00}$ mode and a higher order transverse mode (see also text).}
   \label{fig:omegeff2}
\end{figure}

This fact suggests that it should be possible to observe directly the frequency shift caused by the nonzero second order derivative $\partial^2 \omega/\partial z_0^2$, which is absent in Fabry-Perot cavities with a vibrating micromirror. Eq.~(\ref{omegeff}) suggests that such an effect should be visible around optical resonance $\Delta\simeq 0$, where the frequency shift caused by the radiation pressure interaction [the second fractional term in Eq.~(\ref{omegeff})] is negligible. This is confirmed by the data shown in Fig.~\ref{fig:omegeff2}, where the mechanical frequency shift $\Omega_m^{\rm eff}-\Omega_m$ is plotted for a membrane position $z_0$ ranging between a node and an antinode of the cavity field. Black dots are the experimental data, the red full line refers to the theoretical prediction of Eq.~(\ref{omegeff}) with the same parameters of Fig.~\ref{fig:spectra} except that ${\mathcal P}=76$ $\mu$W, and $\Delta = 0$.
It is evident that the second order correction, even though small, is appreciable, and that including the quadratic term in the dependence of the cavity frequency upon the membrane deformation $\hat{q}$ is necessary for a proper description of the experimental results. The experimental data depart from the theory prediction in a narrow $z_0$ interval which corresponds to the presence of an avoided crossing between the driven TEM$_{00}$ mode and a higher-order mode, which is responsible for an appreciable modification of the optomechanical coupling $G$~\cite{Karuza2011,Sankey2010}.

Measuring the position spectrum allows to measure also the effective mean thermal phonon number $n_{\rm eff}$, e.g. the effective temperature $T_{\rm eff}$, of the vibrational mode of interest. In fact, one has in general~\cite{marquardt,wilson-rae,Genes08,dantan08}
\begin{equation} \label{meanenergy}
\hbar \Omega _{m}\left( n_{\rm eff}+\frac{1}{2}\right) \equiv \frac{\hbar \Omega _{m}}{2}\left[ \left\langle \delta \hat{q}^{2}\right\rangle +\left\langle
\delta \hat{p}^{2}\right\rangle \right],
\end{equation}
where the position and momentum variances can be expressed in terms of the noise spectrum $S_{q}(\omega )$ as 
\begin{equation}
\left\langle \delta \hat{q}^{2}\right\rangle =\int_{-\infty }^{\infty }\frac{%
d\omega }{2\pi }S_{q}(\omega ),\;\;\;\;\left\langle \delta \hat{
p}^{2}\right\rangle =\int_{-\infty }^{\infty }\frac{d\omega }{2\pi }\frac{%
\omega ^{2}}{\Omega _{m}^{2}}S_{q}(\omega ).  \label{spectra}
\end{equation}%
However, as shown in Refs.~\cite{Genes08,dantan08}, energy equipartition $\left\langle \delta \hat{p}^{2}\right\rangle \simeq \left\langle \delta \hat{q}^{2}\right\rangle$ holds in a large parameter regime implying
\begin{equation}
\left\langle \delta \hat{q}^{2}\right\rangle \simeq n_{\rm eff}+\frac{1}{2} \equiv \frac{k_B T_{\rm eff}}{\hbar \Omega_m},
\label{teff}
\end{equation}
where the latter definition can be applied far from the quantum regime ($n_{\rm eff} \sim 1$), pertaining to our experimental situation. Therefore $T_{\rm eff}$ can be obtained evaluating the area below the mechanical resonance peak after subtraction of the flat noise floor due to detection noise (see Eq.~(\ref{detect3})).

However, one has two further ways of inferring $T_{\rm eff}$ from the measured position noise spectrum. A first way is provided by the values of $\gamma_m^{\rm eff}/\gamma_m$ of Fig.~\ref{fig:dampeff}: in fact from a thermodynamical point of view, due to radiation pressure cooling, the vibrational mode passes from a thermal environment characterized by a Langevin force with strength proportional to $\gamma_m T$ to an effective one with a Langevin force of the same strength, which is however proportional to $\gamma_m^{\rm eff}T_{\rm eff}$, so that~\cite{Mancini1998,Cohadon1999}
\begin{equation}
T_{\rm eff}^{\gamma}\equiv T\frac{\gamma_m}{\gamma_m^{\rm eff}} .
\end{equation}
A second temperature measurement is provided by the height of the resonant peak of the calibrated spectrum $S_x(\omega)$. In fact, the extremely small value of $\Gamma$, and the fact that our experiment is carried out at room temperature $T=295$ K and in the weak coupling limit (it is $G \simeq -0.01 \Omega_m$ at $\Delta= \Omega_m$ in Figs.~\ref{fig:dampeff}-\ref{fig:omegeff}) implies that, as it is common in most optomechanical schemes, both the radiation pressure contribution $S_{\rm rp}(\omega)$ and the absorption contribution $S_{\rm abs}(\omega)$ are negligible with respect to that of thermal noise. Therefore one can safely assume
\begin{equation} \label{posspe2}
S_{q}(\omega )\simeq \frac{2\gamma _{m}k_B T}{\hbar \Omega_m} \left|\chi _{\rm eff}(\omega )\right|^{2}.
\end{equation}
Using the fact that at the resonant peak $\omega=\Omega_m^{\rm eff}$ it is $\left|\chi _{\rm eff}\right|^{2}=\left[\Omega_m/\Omega_m^{\rm eff} \gamma_m^{\rm eff}\right]^2$, and using again $\gamma_m^{\rm eff}T_{\rm eff}=\gamma_m T$, one can write from Eq.~(\ref{posspe2})
\begin{equation} \label{posspe3}
S_{q}(\omega_m^{\rm eff} )=\frac{2k_B \Omega_m }{\hbar \gamma_m (\Omega_m^{\rm eff})^2 T} \left[T_{\rm eff}^p\right]^2,
\end{equation}
providing the definition of a ``peak'' temperature $T_{\rm eff}^p$.

\begin{figure}[h]
   \centering
   \includegraphics[width=.8\textwidth]{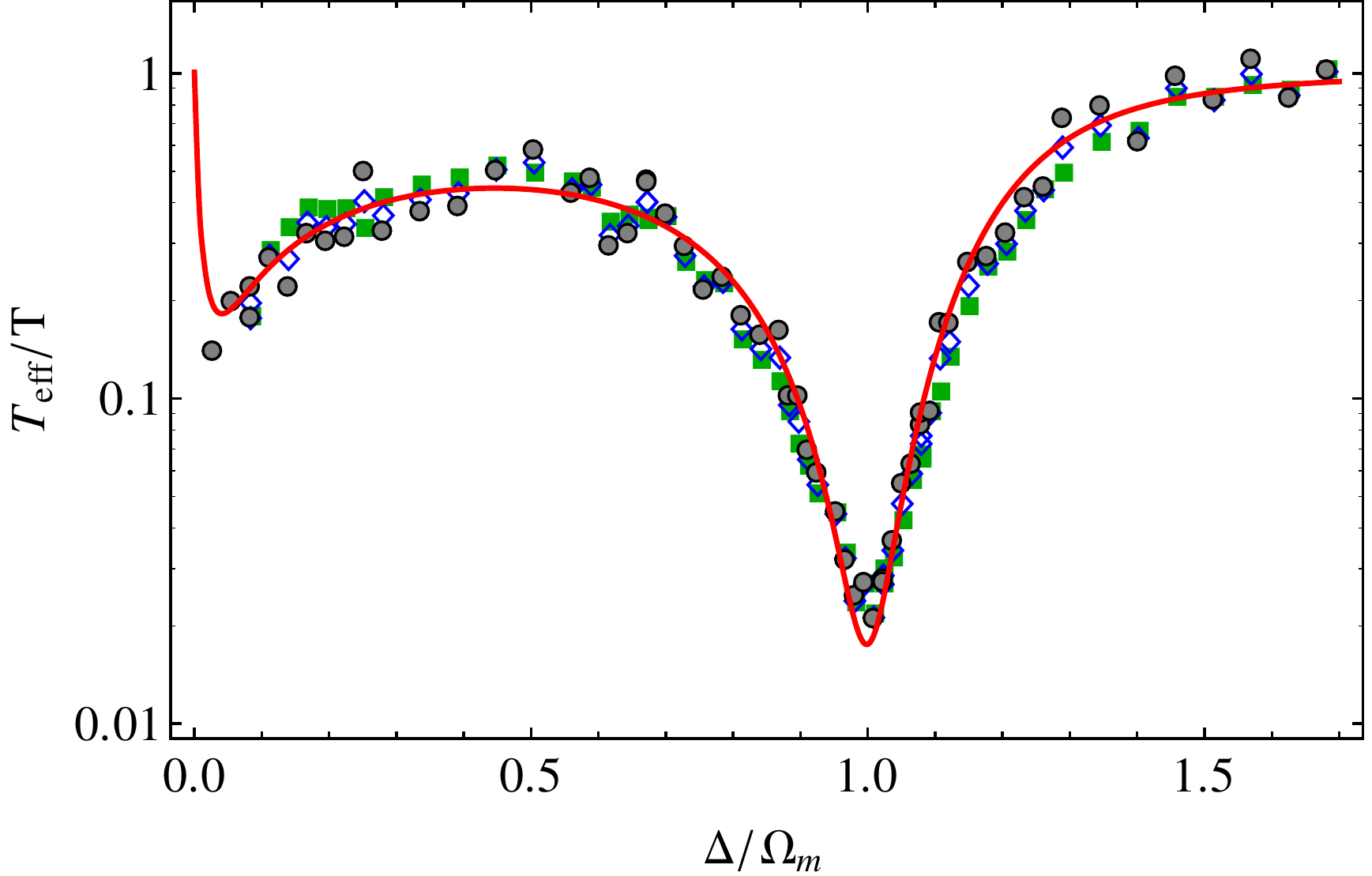}
   \caption{The three different estimates of the vibrational mode temperature, $T_{\rm eff}^{\gamma}$, $T_{\rm eff}^p$ and $T_{\rm eff}^{\rm area}$ vs the scaled cavity-driving detuning $\Delta/\Omega_m$. The other parameters are as in Fig.~\protect\ref{fig:spectra}. Circles correspond to $T_{\rm eff}^{\gamma}$, diamonds to $T_{\rm eff}^p$, and squares to $T_{\rm eff}^{\rm area}$.}
   \label{fig:temp0}
\end{figure}

\begin{figure}[h]
   \centering
   \includegraphics[width=.8\textwidth]{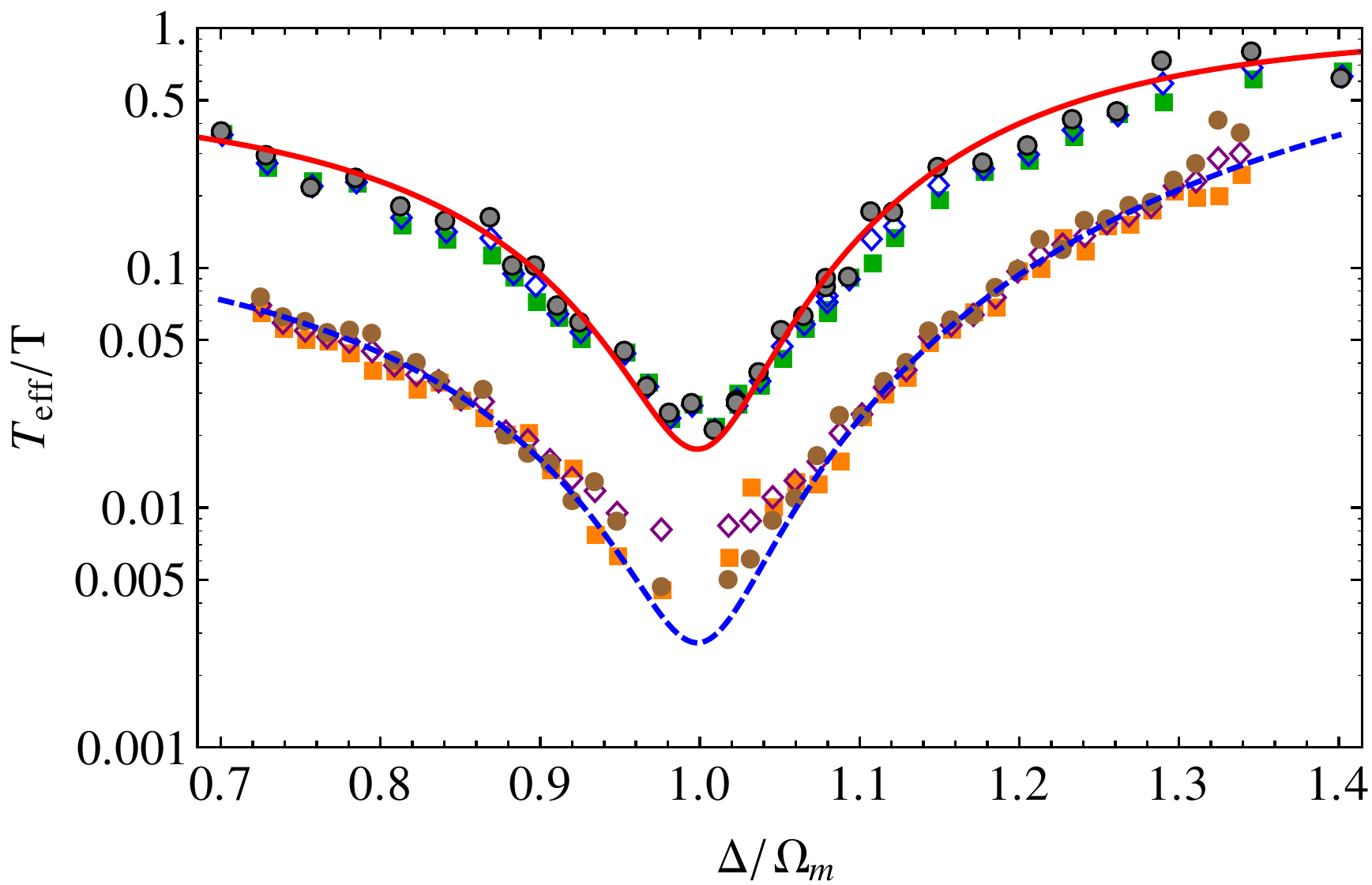}
   \caption{The three different estimates of the vibrational mode temperature, $T_{\rm eff}^{\gamma}$, $T_{\rm eff}^p$ and $T_{\rm eff}^{\rm area}$ vs the scaled cavity-driving detuning $\Delta/\Omega_m$, around resonance with the red sideband of the cavity mode frequency, $\Delta/\Omega_m =1$. The upper curve refers to the small coupling regime of Fig.~\protect\ref{fig:spectra}, while the lower curve refers to a different situation with larger coupling ($G \simeq -0.031 \Omega_m$ at $\Delta =\Omega_m$), obtained with the membrane at $15$ nm distance from a field node, and larger driving input power ${\mathcal P}=1.60$ mW. Circles correspond to $T_{\rm eff}^{\gamma}$, diamonds to $T_{\rm eff}^p$, and squares to $T_{\rm eff}^{\rm area}$.}
   \label{fig:temp}
\end{figure}

The three different estimates for the effective vibrational mode temperature, $T_{\rm eff}^{\gamma}$, $T_{\rm eff}^p$, and the one associated with Eq.~(\ref{teff}), $T_{\rm eff}^{\rm area}$, are plotted in Fig.~\ref{fig:temp0} versus the scaled cavity-driving detuning $\Delta/\Omega_m$, for the same set of parameters of Figs.~2-4. The three different temperature measurements agree with each other and with the theoretical prediction. A subset of these data around the resonant condition $\Delta/\Omega_m =1$ is then compared with the corresponding data with a larger optomechanical coupling in Fig.~\ref{fig:temp}. The upper curve refer to the small coupling regime of Fig.~\ref{fig:temp0}, while the lower curve, showing better cooling, is obtained in a different experimental condition, with the membrane at $15$ nm distance from a field node (larger $\partial \omega/\partial z_0$) and a driving input power ${\mathcal P}=1.60$ mW, which corresponds to the larger coupling $G \simeq -0.031 \Omega_m$ (at $\Delta =\Omega_m$). In this second case, the vibrational mode is cooled down by a factor $\sim 350$. Moreover, the three different estimates of the effective temperature are again consistent between them and with the theoretical prediction (full line in Fig.~\ref{fig:temp}).

\section{Concluding remarks}

We have performed an experimental study of the optomechanical device formed by a vibrating $50$ nm-thick SiN membrane placed within a high-finesse optical Fabry-Perot cavity at room temperature. We have studied in particular how the radiation pressure interaction with the driven cavity mode modifies the mechanical susceptibility of the vibrational mode of interest due to the back-action of the detuned optical field. The measured mechanical frequency shift and the modified mechanical damping as a function of detuning and optomechanical coupling are well reproduced by a Langevin equation description of the system~\cite{Biancofiore2011}. The increase of damping is equivalent to cooling of the vibrational mode, and we demonstrate a decrease in temperature by a factor $\sim 350$. We also observe a mechanical frequency shift which is not associated with the standard optical spring effect, but that can be explained only taking into account the quadratic term in the parametric dependence of the cavity frequency upon the effective membrane position. This second order term, which is usually negligible in most optomechanical devices, cannot be neglected in a proper description of the present membrane-in-the-middle setup.

\section{Acknowledgments}

This work has been supported by the European Commission (FP-7 FET-Open project MINOS).

                                                                                            %



\end{document}